# Wind Speed Prediction using Deep Ensemble Learning with a Jet-like Architecture

Aqsa Saeed Qureshi[1], Asifullah Khan*[1,2], and Muhammad Waleed Khan[1]

[1]Department of Computer and Information Sciences, Pakistan Institute of Engineering and Applied Sciences, Nilore-45650, Islamabad;

[2]Centre for Mathematical Sciences, Pakistan Institute of Engineering and Applied Sciences, Nilore-45650, Islamabad; asif@pieas.edu.pk

**Abstract**

The wind is one of the most increasingly used renewable energy resources. Accurate and reliable forecast of wind speed is necessary for efficient power production; however, it is not an easy task because it depends upon meteorological features of the surrounding region. Deep learning is extensively used these days for performing feature extraction. It has also been observed that the integration of several learning models, known as ensemble learning, generally gives better performance compared to a single model. The design of wings, tail, and nose of a jet improves the aerodynamics resulting in a smooth and controlled flight of the jet against the variations of the air currents. Inspired by the shape and working of a jet, a novel Deep Ensemble Learning using Jet-like Architecture (DEL-Jet) technique is proposed to enhance the diversity and robustness of a learning system against the variations in the input space. The diverse feature spaces of the base-regressors are exploited using the jet-like ensemble architecture. Two Convolutional Neural Networks (as jet wings) and one deep Auto-Encoder (as jet tail) are used to extract the diverse feature spaces from the input data. After that, nonlinear PCA (as jet main body) is employed to reduce the dimensionality of extracted feature space. Finally, both the reduced and the original feature spaces are exploited to train the meta-regressor (as jet nose) for forecasting the wind speed. The performance of the proposed DEL-Jet technique is evaluated for ten independent runs and shows that the deep and jet-like architecture helps in improving the robustness and generalization of the learning system.

## 1. Introduction

Renewable energy resources have got significant attention in the past few years. Out of all forms of renewable energy, the wind is an increasingly utilized source for the generation of electricity. Wind turbines convert kinetic energy of the wind to mechanical energy, and the connected generators convert that mechanical energy into electrical energy. However, efficient usage of wind for power production requires accurate, reliable, and timely prediction of the wind speed. Because of the dependency of the wind on meteorological features of the surrounding region, the prediction of wind speed is a challenging task. One of the most important techniques for feature extraction is deep learning. Furthermore, the integration of different learning models, called ensemble learning, generally gives much better performance in terms of prediction than a single learning model.

Different techniques related to Deep Neural Networks (DNNs) and ensemble learning have been reported in the literature. In DNNs, data representation is learned automatically in such a way that each higher-level layer extracts better feature representation. Different types of DNNs have been employed for different applications [1] [2] [3] [4] [5] [6][7] [8] [9] [10] [11] [12][13] [14]. Wang et al. [1] have reported a 3-D feature learning methodology that utilizes convolutional Auto-Encoder (AE) along with extreme machine learning. Another AE based shape retrieval system has been reported by Zhu et al. [2]. In Zhu's technique, after performing 3D to 2D transformation, AE is used for learning useful features. Similarly, a fault diagnostic system for roller bearing has been reported by Xu et al. [3]. In Xu's technique, two phases are involved. In the first step, denoising AE is used to reduce the dimensionality of feature space. After that, the Gath-Geva algorithm is used to diagnose the fault in the second phase. Lian et al. [4] have used AE for finding a multi-label relationship. For experimental purposes, six different datasets with different levels of noise are used to check the effectiveness of Lian's technique. Wang et al. [5] have used Convolutional Neural Network (CNN) for forecasting photovoltaic power for dataset collected from photovoltaic farms located in China. In another approach, Sun et al. [6] have used CNN for the development of breast cancer diagnostic system. During the development of the diagnostic system, both labeled and unlabeled data is used during training. Similarly, Qiu et al. [8] have developed deep learning-based load demand forecasting system. To improve the generalization performance of CNN, Sun et al. [9] have introduced companion objective function. In Sun's technique, during the experiment, the performance of regularized CNN is checked on different versions of the CIFAR dataset. Similarly, for brain image segmentation, Zhang et al. [10] have used deep CNN for segmenting different iso intensive stages within the brain. During the experiment, different MR images are used to check the performance of Zang's technique. Xuebin et al. [11] have proposed a Deep Belief Network (DBN) based approach for the diagnosis of the fault within the cable. Another harmony search approach has been reported by Papa et al. [15] to fine-tune the parameters of DBN. In another adaptive self-taught learning-based IDS, pre-trained deep sparse AE on the regression related dataset is modified and tuned accordingly, to improve the performance of the network on IDS related dataset [13]. Similarly, deep CNN based technique, which is capable of handling class skewness problems in breast cancer related histopathological images, has been reported by Wahab et al. [14].

Ensemble learning techniques have also been reported in literature in order to obtain robust learning systems [16] [17] [18] [19] [20] [21]. Normaly, generalization capability of an ensemble learner is better than that of individual base-learners. Similarly Huang et al., [16] have used Support Vector Machine (SVM) based ensemble learning approach which combines semantic, spectral as well as structural features of remotely sensed imagery. Huang's technique shows improved performance, when comparison is made with probabilistic and voting based techniques. Ahmed et al., [17] have used Transfer Learning (TL) based ensemble learning approach for developing efficient churn prediction system and achieved an accuracy of 68.2 and 75.4% on cell2cell and orange dataset respectively. Another TL and CNN based technique for detecting mitosis in breast cancer histopathological images has been reported by Wahab et al. [18]. In Wahab's technique, use of TL not only reduces the training time, but also results in an accuracy of 76% and AUC-PR of 0.713.

The proposed DEL-Jet is an ensemble-based learning approach that is evaluated on the Wind Speed (WS) dataset. WS forecasting strategies are generally categorized into five different methods. The first method is called the persistence method, which is used for forecasting the ultra-short forecasting of WS. The persistence method assumes that future WS is the same as the current value of wind forecasting holds. The second method is a physical method that is based on a statistical approach. The physical method requires historical data to model complicated mathematical relationships for reliable long term WS forecasting. The third method of WS forecasting is a statistical approach in which historical data is used for finding a statistical model for reliable short-term forecasting of WS. The fourth method is related to forecasting strategies based on Artificial Intelligence (AI). In this technique, instead of any predefined statistical or complex mathematical models, AI-based technique such as Artificial Neural Networks (ANNs) are utilized for finding a complicated nonlinear relationship between the input variables and target. The fifth method relates to a hybrid technique that combines more than one method for forecasting WS.

Various researchers have proposed interesting approaches for reliable forecasting of WS and wind power. Torres et al. [22] have used Autoregressive-Moving-Average for the hourly prediction of WS. In another technique, Li et al. [23] have proposed an efficient WS prediction strategy that is based on different NN and Bayesian models. Similarly, Wang et al. [24] have suggested a hybrid technique in which quantile regression, along with Wavelet Transform (WT) and DBN are used for forecasting WS. In Wang's work, the dataset has been collected from the wind farms of Australia and China. Wan et al. [25] have proposed an Extreme-Machine-Learning (EML) strategy along with a bootstrapping based model selection strategy for reliable forecasting of wind power. Similarly, various ensemble-based efficient wind speed and power prediction techniques have also been reported in the literature [26–31]. Kretzschmar et al., [32] have used NN for WS and wind-gust forecasting. Another technique utilizes deep NN-based Stacked AE and stacked denoising AE for WS forecasting [33]. Different researchers have also applied SVM [34][35] for wind speed and power forecasting. Another technique that uses three hybrid methods based on ANN and wavelet packet has been reported by Lu et al. [36]. Different ensemble-based learning strategies have also been reported in the literature for reliable forecasting of WP and speed. Zameer et al. [20] have proposed an ensemble-based technique for short term forecasting of wind power. Another DNN-based ensemble learning approach has been reported by Qureshi et al. [19]. In their work, the idea of TL is applied across the base-regressors trained on wind farms located in different regions. Another adaptive TL based wind power prediction strategy has been reported in which the idea of TL is applied not only across different wind farms but also within the same wind farm [37]. A short term-based wind power forecasting technique has been reported by Men et al. [21] in which different mixture density-based NNs are used for forecasting wind speed and power. In general, ensemble-based techniques exploit the decision/feature space of individual learners. Whereby, the final decision is generally made either through voting techniques or by using a Meta learner. However, in the proposed DEL-JET based ensemble learning approach, the feature spaces of the individual learners are first transformed through nonlinear PCA (working as the main body of the jet architecture). This selected and reduced feature space is then combined with the original feature space. The combined feature space is provided to the Meta learner (a combiner ensemble working as the nose of the jet architecture).

The rest of the paper is organized as followed. In section 2, the proposed technique is discussed. Section 3 discusses the details related to the implementation of the proposed technique. Finally, results and conclusions are drawn in section 4 and section 5, respectively.

## 2. Proposed DEL-Jet Technique

The proposed DEL-Jet is a jet-based ensemble learning approach inspired by the architecture of jet plane, in which the main parts are wings, tail, body, and nose. Each part performs different functions that ultimately help in the smooth and desired flight of the jet. Wings are helpful in take-off and the lift to hold the jet in the air. The tail provides stability and thus helps the jet to keep flying straight. The proposed DEL-Jet technique is comprised of three phases. In the first phase, input features are provided to two separate deep CNN and one auto-encoder. After that, all the extracted features from the trained models are concatenated with input features to form transformed features space. In the second phase, nonlinear PCA is used to reduce the dimensionality of the concatenated feature space that is obtained from the first phase. In the third phase, reduced features space is combined with the original feature space, and provided as an input to train multilayer perceptron (MLP) to predict the final wind speed. Block diagram of the proposed method is shown in Figure 1. In the proposed approach, the application of two CNNs and one auto-encoder help to improve the feature representation just like the wings and tail of a jet help in better flight. However, reduced feature space in the second phase of the proposed DEL-Jet technique depicts the main body of a jet, which is the actual part to be moved in the forward direction during a flight. Reduced feature space for the third phase reduces the computational cost of the meta-regressor. The meta-regressor used in the proposed technique, by mimicking the noise of a jet, predicts the final wind speed based on feature space extracted from base regressors.

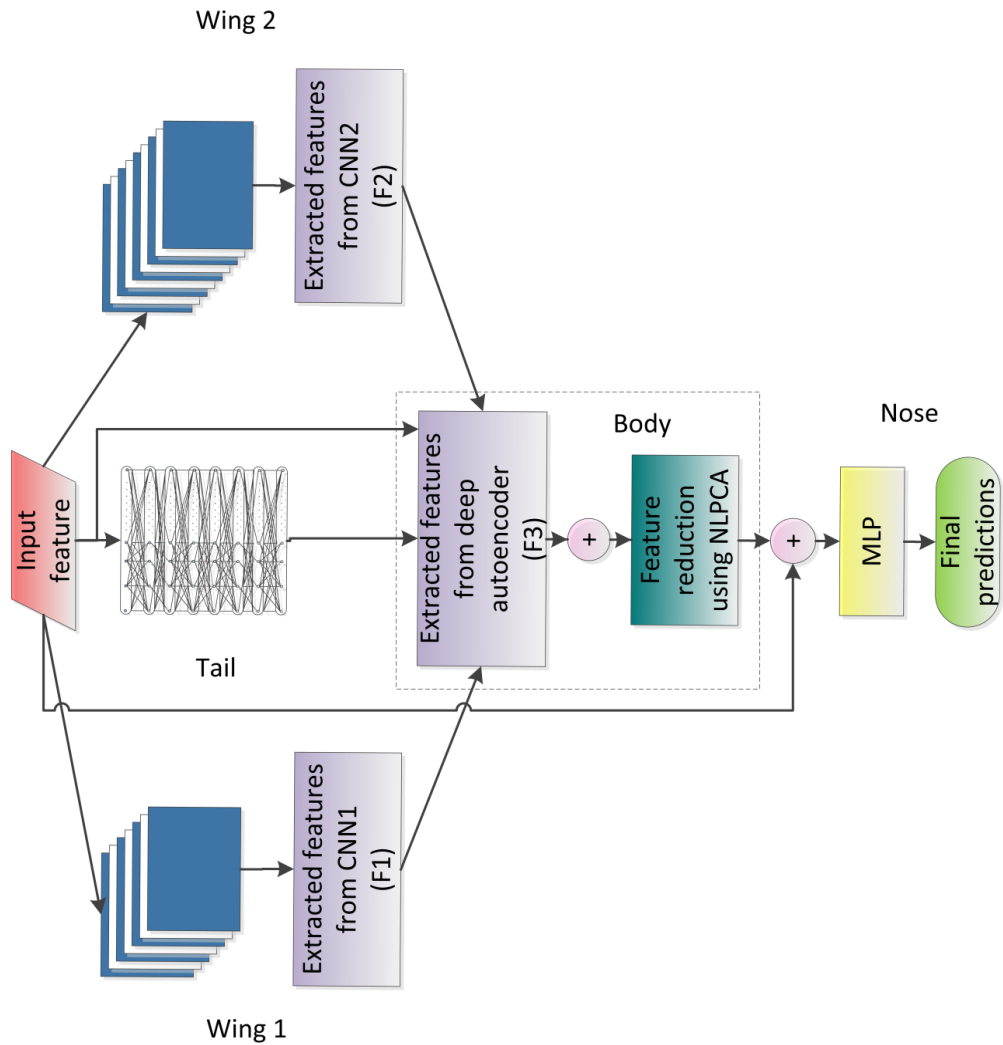

Figure 1: Proposed DEL-Jet technique

**2.1. Wind Speed Data Set**

The data set used in the proposed technique has been collected from a wind farm located in Jhimpir (Pakistan). The wind farm with an area of almost 1283 acres was constructed in October 2012, and a total of 134 million USD were spent in its development. In December 2012, the wind farm started producing electricity on the commercial scale, having a capacity of 49.5 MW.

In this work, the task is to forecast the speed of wind after every ten minutes. So, between a time interval of 10 minutes, average and standard deviation of different features are computed and taken as input against wind speed to be predicted. In the proposed DEL-Jet technique, eight features are used as input features. Four features are related to the mean and standard deviation of wind direction at the heights of 78.5 m and 28.5 m above the ground. Two features keep the information related to the mean of temperature at the heights of 5 m and 80 m from the ground surface. The remaining two features contain the values of humidity and pressure.

The data set contains the wind speed at the heights of 81.5 m, 80 m, 60 m, and 10 m above the ground surface, and the task is to find out the average value of wind speed at different heights after every 10 minutes. Wind speed prediction related data that is used in the proposed DEL-Jet technique is of the time series nature, and each data sample depends on the wind speed and input features of past intervals. So, in order to capture the information related to history, input features against power to be predicted at a time are concatenated with input features and respective predicted wind speed at time t-10, t-20, t-30, t-40, t-50, t-60, t-70 time. Figure 2 shows the way past information is incorporated into the data set used in the proposed DEL-Jet technique.

| F1(t-70) | S1(t-70) | F1(t-60) | S2(t-60) | F1(t-50) | S1(t-50) | F1(t-40) | S1(t-40) | F1(t-30) | S1(t-30) | F1(t-20) | S1(t-20) | F1(t-10) | S1(t-10) | F1(t) | **S1(t)** |
|---|---|---|---|---|---|---|---|---|---|---|---|---|---|---|---|
| F2(t-70) | S2(t-70) | F2(t-60) | S2(t-60) | F2(t-50) | S2(t-50) | F2(t-40) | S2(t-40) | F2(t-30) | S2(t-30) | F2(t-20) | S2(t-20) | F2(t-10) | S2(t-10) | F2(t) | **S2(t)** |
| F3(t-70) | S3(t-70) | F3(t-60) | S3(t-60) | F3(t-50) | S3(t-50) | F3(t-40) | S3(t-40) | F3(t-30) | S3(t-30) | F3(t-20) | S3(t-20) | F3(t-10) | S3(t-10) | F3(t) | **S3(t)** |
| F4(t-70) | S4(t-70) | F4(t-60) | S4(t-60) | F4(t-50) | S4(t-50) | F4(t-40) | S4(t-40) | F4(t-30) | S4(t-30) | F4(t-20) | S4(t-20) | F4(t-10) | S4(t-10) | F4(t) | **S4(t)** |
| ⋮ | ⋮ | ⋮ | ⋮ | ⋮ | ⋮ | ⋮ | ⋮ | ⋮ | ⋮ | ⋮ | ⋮ | ⋮ | ⋮ | ⋮ | ⋮ |
| FN(t-70) | SN(t-70) | FN(t-60) | SN(t-60) | FN(t-50) | SN(t-50) | FN(t-40) | SN(t-40) | FN(t-30) | SN(t-30) | FN(t-20) | SN(t-20) | FN(t-10) | SN(t-10) | FN(t) | **SN(t)** |

Figure 2: Feature space of data set used in the proposed DEL-Jet technique

### 2.1.1. Division of Data:

The collected data is divided in the following way to test the proposed DEL-Jet technique for the prediction of wind speed. Data for the first eight months (66.67 %) is used during the training phase. Data for the next two months (16.67 %) is used to validate the parameters of all of the regressors used in the proposed technique. Finally, data for the last two months (16.67 %) is used as test data.

### 2.2. Feature Extraction

Since the proposed approach is a jet-like approach based on deep ensemble learning, therefore multiple base regressors act as wings and tail. Base-regressors help to provide improved feature space and boost the performance of meta-regressors. In the proposed DEL-Jet technique, two separate deep CNNs and one sparse auto-encoder are used as base-regressors.

### 2.2.1. CNN as Feature Extractor:

CNN is a specialized type of NN that shows excellent performance on data sets having grid-like topology (like images or time series data). In CNN, the convolution operation is used instead of general matrix multiplications. A convolutional layer is considered as a fundamental building block of a CNN. Convolutional operation behaves like a filter to any input and generates activations or feature maps as output. Different convolution operations simultaneously and automatically learn different filters. The convolutional operator is considered as a linear operation in which weights of kernels are multiplied with a specific portion of input data in order to generate single-valued output. Since the filter is convolved multiple times with input, therefore output generated after each layer of convolution is in the form of the two-dimensional feature map. During the training of a CNN, each layer of CNN extracts different features, i.e., lower-level layers extract lower-level features, whereas higher-level layers extract higher-level features. At each layer of convolution, different filters extract the feature maps differently. After the generation of feature maps, usage of the pooling layer helps to make the whole CNN invariant to small input changes. The pooling layer also helps to reduce the number of parameters. In the proposed method, two deep CNNs are used as base-regressor. Since the dataset used in this study follows grid-like topology, therefore CNNs help to extract improved or diverse feature space.

### 2.2.2. Deep sparse AE as Feature Extractor:

An AE is a type of NN that tries to reconstruct the input as its output. When training of an AE is done in such a way that it only tries to reduce Mean Squared Error during training, then this leads to overfitting within the network. The sparsity penalty added within the loss function helps to avoid overfitting during training, reduces the complexity of the trained network, and enables generalized and statistical feature learning from the data. In the proposed DEL-Jet technique, one sparse AE along with two CNNs extract diverse feature space that ultimately helps to improve the performance of a meta-regressor.

### 2.2.3. Feature processing Using Nonlinear PCA Approach

Nonlinear PCA, a generalized form of PCA, is a dimensionality reduction technique. Dimensionality reduction using nonlinear PCA can be achieved using AE in which during training phase, identity mapping is performed, and a smaller number of neurons are set as hidden layers. After training of the AE, the output from the hidden layer extracts reduced feature space from high dimensional input feature space. In the proposed DEL-Jet technique, nonlinear PCA is used to reduce the feature space that is formed by combining the features extracted from all of the three base-regressors.

### 2.3. Final Wind Speed Prediction

MLP is considered as a meta-regressor in the proposed JEL-WSP technique because it predicts the final wind speed based on feature space extracted from base-regressors. MLP performs a nonlinear transformation of the input feature space and predicts the final wind speed using output neurons.

## 3. Implementation Details:

All the simulations related to the proposed DEL-Jet technique have been performed on a desktop computer, having a 16-bit Windows operating system and 16 GB RAM. Whereas, Matlab 2016(a) has been used as a programming platform.

### 3.1. Parameter Setting

Tables 1 and 2 show the parameter setting of the two CNNs and one deep sparse AE used as wings and tail, respectively, in the proposed DEL-Jet technique. Nonlinear PCA serves as a dimensionality reduction technique and extracts 200 features from the hidden layer of the auto-encoder. In the end, MLP having 26 numbers of neurons predicts the final wind speed.

Table 1: Parameter of two CNN based base-regressors used in proposed DEL-Jet technique

| Parameter | CNN1 | CNN2 |
|---|---|---|
| Number of layers | 10 | 12 |
| Number of epochs | 5 | 50 |
| Initial learning rate | 0.001 | 0.001 |
| Learning algorithm | SGD with momentum | SGD with momentum |

Table 2: Parameter of deep AE based base-regressor used in proposed DEL-Jet technique

| Layer | No of Neurons | Maximum Epoch | L2 Weight Regularization | Sparsity Regularization | Sparsity Proportion |
|---|---|---|---|---|---|
| 1 | 50 | 50 | 0.00003 | 4 | 0.15 |
| 2 | 100 | 80 | 0.00001 | 3 | 0.1 |
| 3 | 100 | 200 | 0.00002 | 4 | 0.1 |
| 4 | 175 | 75 | 0.00001 | 3 | 0.1 |
| 5 | 50 | 50 | 0.00001 | 4 | 0.1 |
| 6 | 100 | 250 | 0.00001 | 5 | 0.1 |
| 7 | 55 | 100 | 0.00002 | 4 | 0.1 |
| 8 | 70 | 200 | 0.00002 | 4 | 0.1 |

### 3.2. Evaluation Details Related to the Proposed DEL-Jet Method

In evaluating the performance of the proposed technique, Root Mean Square Error (RMSE), mean absolute error (MAE), and standard deviation of error (SDE) are considered as evaluation measures. Equations 1, 2, and 3 show mathematical expressions of the three evaluation measures.

$$RMSE = \sqrt{\frac{1}{m}\sum_{i=1}^{m}(WS_{actual_i} - WS_{predicted_i})} \qquad (1)$$

$$MAE = \frac{1}{m}\sum_{i=1}^{m}|WS_{actual_i} - WS_{predicted_i}| \quad (2)$$

$$SDE = \sqrt{\frac{1}{m}\sum_{i=1}^{m}(Error_i - Mean\_Error_i)^2} \quad (3)$$

In the above equations, $WS_{actual_i}$ and $WS_{predicted_i}$ represent actual and predicted wind speed against the $i^{th}$ data point. Whereas, $Error$ and $Mean\_Error$ show absolute and mean value of error between the predicted and actual wind speed.

## 4. Results and Discussion

In the proposed DEL-Jet like ensemble-based learning approach, all the regressors are trained using RMSE as a loss-function. Before checking the performance of the proposed DEL-Jet technique, performances of simple baseline regressors, i.e., MLP, Support Vector Regression (SVR), and DBN are evaluated on the same data that is reserved as test data for the DEL-Jet technique. Table 3 shows the performance of MLP, SVR, and DBN on the wind speed data set. In Table 3, the performance of the base regressors is expressed in terms of RMSE, MAE, and SDE against ten independent runs. The performance of DBN is found to be best among all of the baseline regressors.

Table 3: Performance of base-line regressors on the data set used in the proposed DEL-Jet technique

| Regressor | DBN | SVR | MLP |
|---|---|---|---|
| RMSE | 0.02581±0.002237 | 0.0527 | 0.02518±0.026712 |
| MAE | 0.01923±0.002391 | 0.0363 | 0.01814±0.001172 |
| SDE | 0.02368±0.000271 | 0.0521 | 0.02502±0.00136 |

### 4.1. Performance of the Proposed Technique in terms of RMSE, MAE, and SDE

After checking the performance of simple baseline regressors, the performance of the proposed DEL-Jet technique is evaluated against ten independent runs. In the proposed technique, two CNN based base-regressors act as wings; deep sparse AE acts as a tail; nonlinear PCA serves as the main body; and MLP act as a nose of the jet ensemble-based approach. Table 4, Table 5, and Table 6 show the performance of wing1, wing2, tail, body, and nose in terms of RMSE, MAE, and SDE. All of the three tables show that performance of two CNNs (wings) and auto-encoder (tail) is excellent, but when extracted feature space from two CNNs and deep sparse AE is further reduced using nonlinear PCA, then the performance is further increased which is clearly depicted in column five of each of the three tables. Whereas, concatenated feature space, when provided to MLP, further increases the performance in terms of all of the evaluation measures. Moreover, the comparison of final wind speed using concatenated feature space is more robust and reliable in comparison to the performance of MLP on original feature space, as shown in Table 3.

Table 4: Performance of the base learners and the proposed DEL-Jet technique in terms of RMSE

|     | Wing1 (CNN1) | Wing2 (CNN2) | Tail (Auto-encoder) | Body (Nonlinear PCA) | Nose (MLP) |
|-----|---|---|---|---|---|
| 1.  | 0.1100 | 0.1103 | 0.1327 | 0.0421 | 0.0237 |
| 2.  | 0.1101 | 0.1102 | 0.0942 | 0.0433 | 0.0249 |
| 3.  | 0.1103 | 0.1102 | 0.1078 | 0.0438 | 0.0248 |
| 4.  | 0.1101 | 0.1100 | 0.1550 | 0.0440 | 0.0243 |
| 5.  | 0.1101 | 0.1102 | 0.1327 | 0.0439 | 0.0241 |
| 6.  | 0.1101 | 0.1102 | 0.1562 | 0.0434 | 0.0242 |
| 7.  | 0.1102 | 0.1102 | 0.1616 | 0.0450 | 0.0250 |
| 8.  | 0.1102 | 0.1100 | 0.1327 | 0.0478 | 0.0265 |
| 9.  | 0.1102 | 0.1103 | 0.1616 | 0.0455 | 0.0242 |
| 10. | 0.1101 | 0.1103 | 0.1616 | 0.0466 | 0.0248 |
|     | 0.11014±0.00008 | 0.11019±0.00010 | 0.13961±0.02276 | 0.04454±0.00161 | 0.02465±0.00073 |

Table 5: Performance of the base learners and the proposed DEL-Jet technique in terms of MAE

|     | Wing1 (CNN1) | Wing2 (CNN2) | Tail (Auto-encoder) | Body (Nonlinear PCA) | Nose (MLP) |
|-----|---|---|---|---|---|
| 1.  | 0.0902 | 0.0904 | 0.0885 | 0.0312 | 0.0172 |
| 2.  | 0.0902 | 0.0904 | 0.0874 | 0.0325 | 0.0181 |
| 3.  | 0.0904 | 0.0904 | 0.0802 | 0.0328 | 0.0180 |
| 4.  | 0.0902 | 0.0901 | 0.0905 | 0.0330 | 0.0175 |
| 5.  | 0.0902 | 0.0904 | 0.0885 | 0.0326 | 0.0174 |
| 6.  | 0.0902 | 0.0903 | 0.0824 | 0.0322 | 0.0174 |
| 7.  | 0.0903 | 0.0904 | 0.0910 | 0.0330 | 0.0181 |
| 8.  | 0.0904 | 0.0901 | 0.0885 | 0.0357 | 0.0193 |
| 9.  | 0.0903 | 0.0904 | 0.0910 | 0.0338 | 0.0174 |
| 10. | 0.0902 | 0.0904 | 0.0910 | 0.0345 | 0.0178 |
|     | 0.09026±0.00008 | 0.09033±0.00012 | 0.08790±0.00356 | 0.03313±0.00120 | 0.01782±0.00058 |

Table 6: Performance of the base learners and the proposed DEL-Jet technique in terms of SDE

|     | Wing1 (CNN1) | Wing2 (CNN2) | Tail (Auto-encoder) | Body (Nonlinear PCA) | Nose (MLP) |
|-----|---|---|---|---|---|
| 1.  | 0.1082 | 0.1082 | 0.1073 | 0.0421 | 0.0237 |
| 2.  | 0.1082 | 0.1082 | 0.1075 | 0.0430 | 0.0249 |
| 3.  | 0.1082 | 0.1082 | 0.0977 | 0.0438 | 0.0248 |
| 4.  | 0.1082 | 0.1082 | 0.1088 | 0.0440 | 0.0243 |
| 5.  | 0.1082 | 0.1082 | 0.1070 | 0.0438 | 0.0240 |
| 6.  | 0.1082 | 0.1082 | 0.0994 | 0.0434 | 0.0242 |
| 7.  | 0.1082 | 0.1082 | 0.1073 | 0.0444 | 0.0249 |
| 8.  | 0.1082 | 0.1082 | 0.1070 | 0.0478 | 0.0264 |
| 9.  | 0.1082 | 0.1082 | 0.1073 | 0.0455 | 0.0242 |
| 10. | 0.1082 | 0.1082 | 0.1073 | 0.0464 | 0.0248 |
|     | 0.1082 | 0.1082 | 0.10566±0.00361 | 0.04442±0.001609 | 0.02462±0.00071 |

### 4.2. Actual and Predicted Wind Speed:

Actual and predicted speed of wind is shown in Figure 3, which clearly depict that the predicted speed of wind is close to actual wind speed.

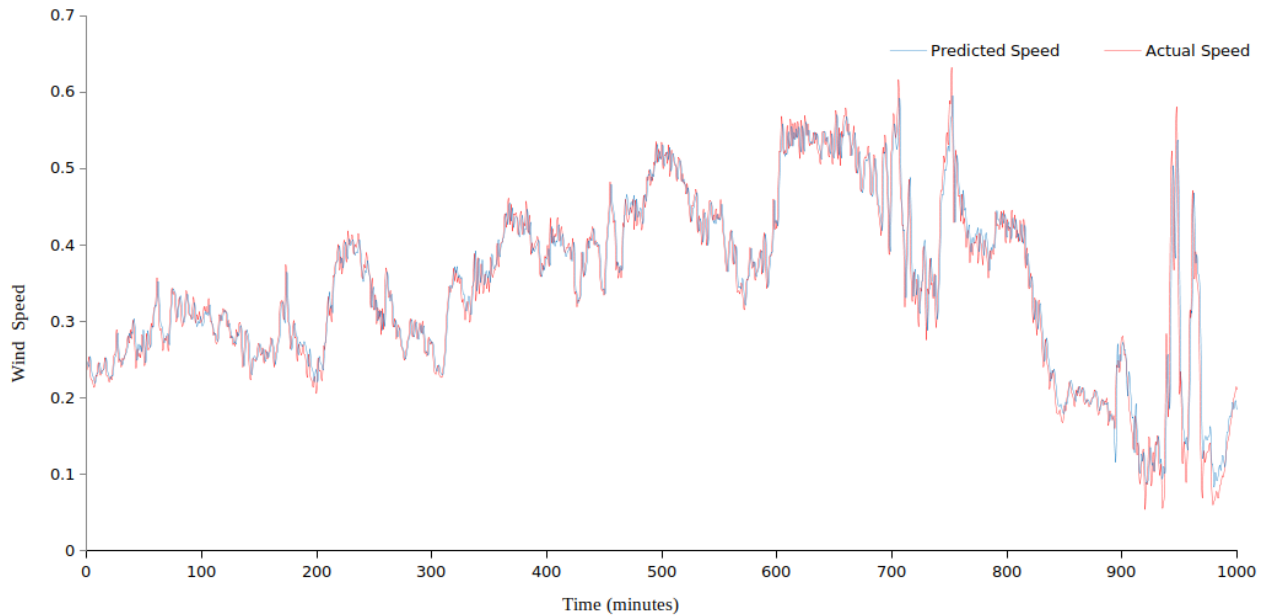

Figure 3: Predicted and Actual speed of wind

### 5. Conclusion

In this paper, a new jet-like deep ensemble learning approach is proposed, which enhances the diversity and robustness of the learning system against variations in the input space. In the proposed DEL-Jet approach, unlike the traditional ensemble learning approaches, the feature spaces of all the base-regressors are concatenated, and then feature transformation is performed using nonlinear PCA. Nonlinear PCA also helps in reducing the dimensionality, and the reduced feature space is then combined with the original feature space. This combination helps in improving the diversity, and thus the performance of the meta-regressor. Experimental results show that the proposed technique is not only robust but also shows excellent performance when compared with the base-regressors.

**References:**

[1] Wang Y, Xie Z, Xu K, Dou Y, Lei Y. Neurocomputing An ef fi cient and effective convolutional auto-encoder extreme learning machine network for 3d feature learning. Neurocomputing 2016;174:988–98. doi:10.1016/j.neucom.2015.10.035.

[2] Zhu Z, Wang X, Bai S, Yao C, Bai X. Neurocomputing Deep Learning Representation using Autoencoder for 3D Shape Retrieval. Neurocomputing 2016;204:41–50. doi:10.1016/j.neucom.2015.08.127.

[3] Xu F, Tse P, Tse YL. Roller bearing fault diagnosis using stacked denoising autoencoder in deep learning and Gath – Geva clustering algorithm without principal component analysis and data label. Appl Soft Comput J 2018;73:898–913. doi:10.1016/j.asoc.2018.09.037.

[4]     Lian S, Liu J, Lu R, Luo X. Captured multi-label relations via joint deep supervised autoencoder. Appl Soft Comput J 2019;74:709–28. doi:10.1016/j.asoc.2018.10.035.

[5]     Wang H, Yi H, Peng J, Wang G, Liu Y, Jiang H, et al. Deterministic and probabilistic forecasting of photovoltaic power based on deep convolutional neural network. Energy Convers Manag 2017;153:409–22. doi:10.1016/j.enconman.2017.10.008.

[6]     Sun W, Tseng TB, Zhang J, Qian W. Computerized Medical Imaging and Graphics Enhancing deep convolutional neural network scheme for breast cancer diagnosis with unlabeled data. Comput Med Imaging Graph 2017;57:4–9. doi:10.1016/j.compmedimag.2016.07.004.

[7]     Khan A, Sohail A, Ali A. A New Channel Boosted Convolutional Neural Network using Transfer Learning. arXiv Prepr arXiv180408528 2018.

[8]     Qiu X, Ren Y, Suganthan PN, Amaratunga GAJ. Empirical Mode Decomposition based Ensemble Deep Learning for Load Demand Time Series Forecasting. Appl Soft Comput 2017;54:246–55. doi:10.1016/j.asoc.2017.01.015.

[9]     Sun W, Su F. A novel companion objective function for regularization of deep convolutional neural networks ☆. Image Vis Comput 2016;60:58–63. doi:10.1016/j.imavis.2016.11.012.

[10]    Zhang W, Li R, Deng H, Wang L, Lin W, Ji S, et al. NeuroImage Deep convolutional neural networks for multi-modality isointense infant brain image segmentation. Neuroimage 2015;108:214–24. doi:10.1016/j.neuroimage.2014.12.061.

[11]    Yizhe Z, Wang M, Gang D, Jun G, Pai W. A cable fault recognition method based on a deep belief network ☆ 2018;71:452–64. doi:10.1016/j.compeleceng.2018.07.043.

[12]    Khan A, Sohail A, Zahoora U, Qureshi AS. A Survey of the Recent Architectures of Deep Convolutional Neural Networks. arXiv Prepr arXiv190106032 2019.

[13]    Saeed A, Asifullah Q. Intrusion detection using deep sparse auto-encoder and self-taught learning. Neural Comput Appl 2019;6. doi:10.1007/s00521-019-04152-6.

[14]    Wahab N, Khan A, Lee YS. Two-phase deep convolutional neural network for reducing class skewness in histopathological images based breast cancer detection. Comput Biol Med 2017;85:86–97. doi:10.1016/j.compbiomed.2017.04.012.

[15]    Paulo J, Scheirer W, Daniel D. Fine-tuning Deep Belief Networks using Harmony Search. Appl Soft Comput J 2016;46:875–85. doi:10.1016/j.asoc.2015.08.043.

[16]    Huang X, Zhang L. An SVM Ensemble Approach Combining Spectral , Structural , and Semantic Features for the Classification of High-Resolution Remotely Sensed Imagery. IEEE Trans Geosci Remote Sens 2013;51:257–72. doi:10.1109/TGRS.2012.2202912.

[17]    Ahmed U, Khan A, Khan SH, Basit A, Haq IU, Lee YS. Transfer Learning and Meta Classification Based Deep Churn Prediction System for Telecom Industry n.d.:1–9. doi:1901.06091.

[18]    Wahab N, Khan A, Lee YS. Transfer learning based deep CNN for segmentation and detection of mitoses in breast cancer histopathological images 2019:216–33. doi:10.1093/jmicro/dfz002.

[19]    Qureshi AS, Khan A, Zameer A, Usman A. Wind power prediction using deep neural network based meta regression and transfer learning. Appl Soft Comput J 2017;58:742–55. doi:10.1016/j.asoc.2017.05.031.

[20]    Zameer A, Arshad J, Khan A, Asif M, Raja Z. Intelligent and robust prediction of short term wind power using genetic programming based ensemble of neural networks. Energy Convers

Manag 2017;134:361–72. doi:10.1016/j.enconman.2016.12.032.

[21] Men Z, Yee E, Lien F-S, Wen D, Chen Y. Short-term wind speed and power forecasting using an ensemble of mixture density neural networks. Renew Energy 2016;87:203–11. doi:10.1016/j.renene.2015.10.014.

[22] Torres JL, Blas M De, Francisco A De, Garcı A. Forecast of hourly average wind speed with ARMA models in Navarre ( Spain ) 2005;79:65–77. doi:10.1016/j.solener.2004.09.013.

[23] Li G, Shi J, Zhou J. Bayesian adaptive combination of short-term wind speed forecasts from neural network models. Renew Energy 2011;36:352–9. doi:10.1016/j.renene.2010.06.049.

[24] Wang HZ, Wang GB, Li GQ, Peng JC, Liu YT. Deep belief network based deterministic and probabilistic wind speed forecasting approach 2016;182:80–93. doi:10.1016/j.apenergy.2016.08.108.

[25] Wan C, Member S, Xu Z, Member S, Pinson P, Member S. Probabilistic Forecasting of Wind Power Generation 2014;29:1033–44.

[26] Zhao J, Guo Z, Su Z, Zhao Z, Xiao X, Liu F. An improved multi-step forecasting model based on WRF ensembles and creative fuzzy systems for wind speed. Appl Energy 2016;162:808–26. doi:10.1016/j.apenergy.2015.10.145.

[27] Wang S, Zhang N, Wu L, Wang Y. Wind speed forecasting based on the hybrid ensemble empirical mode decomposition and GA-BP neural network method. Renew Energy 2016;94:629–36. doi:10.1016/j.renene.2016.03.103.

[28] Wang H, Li G, Wang G, Peng J, Jiang H, Liu Y. Deep learning based ensemble approach for probabilistic wind power forecasting. Appl Energy 2017;188:56–70. doi:10.1016/j.apenergy.2016.11.111.

[29] Taylor JW, Mcsharry PE, Member S, Buizza R. Wind Power Density Forecasting Using Ensemble Predictions and Time Series Models 2009:775–82.

[30] Sloughter JM, Gneiting T, Raftery AE, Loughter JMS, Neiting TG, Aftery AER. Probabilistic Wind Speed Forecasting Using Ensembles and Bayesian Model Averaging Probabilistic Wind Speed Forecasting Using 2012;1459. doi:10.1198/jasa.2009.ap08615.

[31] Liu H, Tian H, Liang X, Li Y. New wind speed forecasting approaches using fast ensemble empirical model decomposition , genetic algorithm , Mind Evolutionary Algorithm and Arti fi cial Neural Networks. Renew Energy 2015;83:1066–75. doi:10.1016/j.renene.2015.06.004.

[32] Neural Network Classifiers for Local Wind Prediction 2004:727–38.

[33] Khodayar M, Kaynak O, Khodayar ME. Rough Deep Neural Architecture for Short-Term Wind Speed Forecasting 2017;13:2770–9.

[34] Zhou J, Shi J, Li G. Fine tuning support vector machines for short-term wind speed forecasting. Energy Convers Manag 2011;52:1990–8. doi:10.1016/j.enconman.2010.11.007.

[35] Zhai Y. WIND SPEED FORECASTING BASED ON SUPPORT VECTOR MACHINE 2007:19–22.

[36] Liu H, Tian H, Pan D, Li Y. Forecasting models for wind speed using wavelet , wavelet packet , time series and Artificial Neural Networks. Appl Energy 2013;107:191–208. doi:10.1016/j.apenergy.2013.02.002.

[37] Qureshi AS, Khan A. Adaptive transfer learning in deep neural networks: Wind power

prediction using knowledge transfer from region to region and between different task domains. Computational Intelligence. 2019;1–26. https://doi.org/10.1111/coin.12236